\documentclass[pre,twocolumn,showpacs]{revtex4}
\usepackage{graphics,graphicx,dcolumn,bm,fleqn,epic,eepic,float}
\usepackage{amssymb,amsmath,multirow,rotate,color}
\usepackage{subfigure}

\def \bxi {\mbox {\boldmath $\xi$}}
\def \bm {\mbox {\boldmath $m$}}
\def \bA {\mbox {\boldmath $A$}}
\def \bB {\mbox {\boldmath $B$}}
\def \bS {\mbox {\boldmath $S$}}
\def \bX {\mbox {\boldmath $X$}}

\begin{document}

\setcounter{page}{0}

\title{\bf Period-two cycles in a feed-forward layered neural network
model with symmetric sequence processing}
\author{F. L. \surname{Metz} and W. K. \surname{Theumann}}
\affiliation{Instituto de F\'\i sica, Universidade Federal do Rio
Grande do Sul, Caixa Postal 15051, 91501-970 Porto Alegre, Brazil}

\date{\today}
\thispagestyle{empty}

\begin{abstract}

The effects of dominant sequential interactions are investigated in
an exactly solvable feed-forward layered neural network model of
binary units and patterns near saturation in which the interaction
consists of a Hebbian part and a symmetric sequential term. Phase
diagrams of stationary states are obtained and a new phase of cyclic
correlated states of period two is found for a weak Hebbian term,
independently of the number of condensed patterns $c$.
\end{abstract}


\maketitle \setcounter{page}{1}

\section{Introduction}

The dynamics and the stationary states of recurrent attractor neural
networks that process sequences of patterns have been studied over
some time and there has been a recent revival of interest near the
storage saturation limit in large networks [1-8]. Either a process
involving a sequence of patterns [3-7] (referred to in this paper as
asymmetric sequence processing), leading to a stationary limit
cycle, or a process involving a pair of sequences, one with patterns
in increasing order and the other one with patterns in decreasing
order (referred to as symmetric sequence processing in what
follows), were considered in those works.

Symmetric sequence processing competing with pattern reconstruction
favored by a Hebbian term of fixed strength has also been considered
\cite{CT94,UHO04}. The ratio $J_{H}/J_{S}$ between the strengths
$J_{H}$ and $J_S$ of the Hebbian term and of the pair of sequences,
respectively, has been restricted to a mostly dominant Hebbian
strength, that is to $1\leq J_{H}/J_{S}\leq\infty$, leading to phase
diagrams which only exhibit fixed-point solutions including
non-trivial correlated attractors \cite{CT94,UHO04}. These are
states that indicate a selectivity in response to a set of
previously learnt patterns in which the correlation coefficients for
the attractors with increasingly distant patterns from a stimulus
are decreasing functions which eventually become vanishingly small.
The sequential part of the interaction induces transitions between
patterns, whereas a sufficiently strong Hebbian term locks the
transitions favoring single-pattern recognition. The case of network
models with a weak Hebbian interaction competing with a dominating
symmetric sequential processing in which $J_{H}/J_{S}<1$ has,
apparently, not been studied before and it is important to find out
what kind of solutions appear in that case and their possible
biological implications.

Indeed, the connection between the information input in a network
of contiguous stimuli in a training sequence of uncorrelated
patterns and correlated delay activity as an output has been of
great interest to explain the results of experimental recordings
of a visual-memory task in the inferotemporal (IT) cortex of
monkeys [8-13], in which correlated states play an essential role.
The results can be interpreted as a connection between persistent
cortex activity and long-term associative memory described by a
fixed-point attractor dynamics, and the early model calculations
that have been done are based on the competition between pattern
reconstruction and symmetric sequence processing
\cite{GTA93,CT94}.

There could be other attractors that may predict further behavior
on visual-memory in the primate IT cortex viewed as a dynamical
system, generating either limit cycles, chaotic or other kinds of
behavior. Earlier model calculations on the competition between
pattern reconstruction and asymmetric sequence processing exhibit
periodic and stationary fixed-point attractors, besides
quasi-stationary states \cite{CS92,MT05}. Motivated by the results
that appeared in the type of experiments on the IT cortex in
monkeys and their interpretation [1,8-13], it would be interesting
to see if there are periodic attractors that appear as correlated
states in a neural network model with the kind of competition
between pattern reconstruction and symmetric sequence processing
that has been studied before in a recurrent network
\cite{GTA93,CT94}, now with a weak Hebbian term. The presence of
periodic attractors would suggest a new kind of persistent
oscillating activity in the IT cortex. The main purpose of this
work is to explore this issue and with that in mind one has to
resort to a dynamical procedure even to detect stationary cyclic
behavior and to determine the emergent properties.

Since the dynamics of fully recurrent neural networks with binary
units is already fairly complicated, and even more so with
graded-response or other more realistic units, we make a drastic
simplification to start with in order to get to the essentials of
our issue. We are mainly interested in finding out if there are
cyclic attractors and in characterizing their properties for a
sizeable critical storage of patterns. Our interest is in the role
of competing interactions that favor either transitions between
patterns or the recognition of specific patterns. To that end we
focus on a simple attractor feed-forward layered network model of
binary units and patterns with a parallel dynamics and without
lateral connections \cite{DKM89}. The model is exactly solvable and
has been extensively used in the past as a model for associative
memory that has all the stationary features of a recurrent network
and it is particularly suited to detect stationary non-equilibrium
states. We show that, despite sequential learning, the procedure
involves in practice a finite number of recursion relations even for
a macroscopically large system. The effect of the lateral
connections is to change the quantitative results of the model.

To make that point clear we construct the phase diagrams that
describe the network behavior for an arbitrary strength of the
Hebbian term in order to check first that we get the same
qualitative behavior as that already found for a recurrent network
in the case of balanced interactions or for a dominant Hebbian
term. Having studied that part, one can be reasonably confident
that the behavior of the layered network for dominant sequential
interaction should also describe the correct qualitative behavior
of a recurrent network. The fully quantitative behavior of that
network is a relevant issue which is beyond the scope of the
present paper.

It will be shown that a phase of cyclic stationary solutions of
period two is obtained, independently of the number of condensed
patterns $c$ with macroscopic overlap with the states of the
network, and that this is a phase of non-trivial correlated states.
These are states that could reflect a new kind of persistent
activity in visual-memory task experiments. On the other hand, the
stability of the cyclic phase is strongly dependent on $c$. These
features distinguish the properties of the present model from those
for asymmetric sequence processing competing with pattern
reconstruction where cycles of period $c$ appear, for arbitrary $c$
\cite{MT05}.

The purpose of studying a simplified model which has the qualitative
features of a recurrent network with more realistic interactions is
that it tells what to look for and what could be relevant in a more
biological minded network. The paper is organized as follows. In
Sec. 2 we present the model and outline the derivation of the
dynamics with the aid of an appendix. We present our main results in
Sec. 3 and conclude with a further discussion in Sec. 4.

\section{The model and the dynamics}

The network model consists of $L$ layers, each containing $N$
binary units (neurons) in states $S_i(l)=\pm 1$, where $i$ denotes
the unit and $l$ the layer. The state $+1$ represents a firing
unit and the state $-1$ a unit at rest. The state of each unit on
a given layer is determined in parallel by the state of all units
on the previous layer, the layer label acting as a time step,
according to the stochastic rule with probability \cite{DKM89}
\begin{eqnarray}
{\rm P}(S_i(l+1)|\bS(l))&=&\frac{\exp[\beta S_i(l+1)h_i(l+1)]}
{2 \cosh[\beta h_i(l+1))]}\,\, , \label{prob} \\
h_i(l+1)&=&\sum^{N}_{j=1}J_{ij}(l)S_j(l) \,\,\,\,,
\label{campodef}
\end{eqnarray}
where $h_i(l+1)$ is the local field at unit $i$ on layer $l+1$ due
to the set of states $\bS(l)$ of all units on layer $l$ and
$J_{ij}(l)$ is the synaptic coupling between unit $j$ on layer $l$
and unit $i$ on layer $l+1$. There is no feedback in the updating
of the units and the first layer has to be set externally in a
given state. The parameter $\beta=T^{-1}$ controls the synaptic
noise such that the dynamics is fully deterministic when
$T\rightarrow 0$ and fully random when $T \rightarrow \infty$.

A macroscopic set of $p=\alpha N$ statistically independent and
identically distributed random patterns $\{ \bxi^{\mu}(l) \}$,\,
$\mu = 1, \dots, p$, with components $\xi_i^{\mu}(l)=\pm 1$ and
probability $\frac{1}{2}$ for either value, are stored on every
layer independently of other layers, according to the learning rule
\begin{equation}
  J_{ij}(l)=\frac{1}{N}\sum_{\mu,\,\rho=1}^p\xi_i^{\mu}(l+1)
  X_{\mu \rho} \xi_j^{\rho}(l)\,\,\,. \label{Jij}
\end{equation}
Thus, there are only interactions between pairs of units on
consecutive layers. Connections between units on more distant
layers, as well as lateral connections between units on the same
layer are excluded. Here, $X_{\mu \rho}$ are the elements of the
matrix
\[  \bX =  \left(\begin{matrix} \bA && 0 \cr 0 && \bB
  \end{matrix} \right ) \,\,,\]
and
\begin{eqnarray}
A_{\mu \rho}&=&\nu \delta_{\mu,\, \rho}+(1-\nu)\,(\delta_{\mu,\,
\rho+1} + \delta_{\mu,\, \rho-1}) \,\,, \\ \nonumber B_{\mu
\rho}&=& b \delta_{\mu,\, \rho}+(1-b)\,(\delta_{\mu,\, \rho+1} +
\delta_{\mu,\, \rho-1}) \,\,, \label{5}
\end{eqnarray}
are the elements of the $c \times c$ and $(p-c) \times (p-c)$ blocks
$\bA$ and $\bB$ responsible for the signal and for the noise in the
local field, respectively, and $c$ is the number of condensed
patterns that yield macroscopic overlaps defined below. The diagonal
two-block interaction matrix reflects the fact that the patterns are
associated in two independent cycles, one for the condensed patterns
($\bxi^{c+1}(l) = \bxi^{1}(l)$) and the other one for the
non-condensed patterns ($\bxi^{p+1}(l) = \bxi^{c+1}(l)$). This
guarantees the applicability of the procedure. The first parts of
$\bA$ and $\bB$ contribute to a Hebbian interaction $J_H$ and their
second parts contribute to the symmetric sequential interaction
$J_S$. The training of the network model may be thought to proceed
in two stages assuming the patterns are numbered in a given order.
In one stage the set of patterns is presented to the network in
random order, every pattern being presented the same number of
times. This builds up the Hebbian part of the learning rule, whereas
the sequential part of the rule takes place as follows in another
stage. The patterns are ordered in two sequences, one sequence in
increasing order in which each pattern is presented with the
following pattern in the sequence, and the other sequence in
decreasing order where every pattern is presented with the previous
pattern.

The crucial parameter is $\nu$ determining the ratio $J_{H}/J_{S}$
of the Hebbian to sequential interaction in the signal term of the
local field. On the other hand, different sequential noise levels
given by $b$ should yield qualitatively similar results as found in
previous works \cite{CT94,MT05}. This does not mean that $b$ is an
irrelevant parameter and we consider this point below. When $b=1$
there is a purely Hebbian noise and for any other $b$ there is a
Hebbian plus sequential noise. In distinction to other works, the
choice ($0\leq\nu,b\leq 1$) enables us to explore the full range of
parameters $\nu$ and $b$.

The macroscopic overlap components $m_{\mu}(l)$ of $O(1)$, between
the configuration $\bS(l)$ and the condensed patterns, are given by
the large-$N$ limit of
\begin{equation}
m^{\mu}_{N}(l)=\frac{1}{N}\sum_{i=1}^{N}\xi_i^{\mu}(l) \langle
S_i(l)\rangle\,\,;\,\, \mu = 1, \dots ,c\,\,, \label{mmacrodef}
\end{equation}
where $\langle \dots \rangle$ denotes a thermal average with
Eq.~(\ref{prob}), whereas the overlaps with the remaining ($p-c$)
non-condensed patterns are $M^{\mu}_{N}(l)=O(1/\sqrt{N})$.

Following the standard procedure for the layered network, one may
write the local field as a sum of a signal and a noise term
$\omega_i(l)$ due to the condensed and the non-condensed patterns,
respectively \cite{DKM89}. The noise follows a Gaussian distribution
with mean zero and a variance $\Delta^{2}(l)$ given by the large-N
limit of
\begin{equation}
\Delta^{2}_N(l)=\sum_{\mu=c+1}^{p} \overline{ \langle \,
Q^{\mu}_{N}(l)^{2} \,
  \rangle} \,,\label{vardef}
\end{equation}
where the overbar denotes the average over {\it all} the patterns,
in which
\begin{equation}
Q^{\mu}_{N}(l) = bM^{\mu}_{N}(l)+(1-b)[M^{\mu-1}_{N}(l) +
M^{\mu+1}_{N}(l)]\,. \label{Qdef}
\end{equation}
The non-condensed overlaps that appear here depend on all patterns
as functions of the full local field. Averages over the
non-condensed patterns can then be performed by integration and we
obtain first the recursion relations for the macroscopic overlaps
$\bm(l)=(m_{1}(l), \dots, m_{c}(l))$
\begin{equation}
\bm(l+1)= \langle\bxi\int
Dz\,\tanh\{\beta[\bxi.\bA\bm(l)+\Delta(l)z]\}\rangle_{\bxi}\,\,\,,
\label{overlaprec}
\end{equation}
where $Dz=e^{-z^{2}/2}dz/\sqrt{2\pi}$ and $\langle \dots
\rangle_{\bxi}$ denotes an explicit average over the condensed
patterns which does not depend on the specific realization of the
patterns.

To obtain a dynamic equation for $\Delta^{2}(l)$ we need recursion
relations not only for the average squared non-condensed overlaps,
$\overline{\langle M^{\mu}_{N}(l)^{2}\rangle}$ and
$\overline{\langle M^{\mu\pm 1}_{N}(l)^{2}\rangle}$, which can be
derived in the usual way \cite{DKM89}, but also for the correlation
of two consecutive overlaps and two next-to-consecutive overlaps,
$\overline{\langle M^{\mu}_{N}(l) M^{\mu\pm 1}_{N}(l)\rangle}$ and
$\overline{\langle M^{\mu-1}_{N}(l) M^{\mu+1}_{N}(l)\rangle}$,
respectively. These generate, in turn, correlations of overlaps
between more distant patterns, which requires to keep track of a
general form \cite{MT05}
\begin{equation}
C_{n}^2(l) = \sum_{\mu=c+1}^{p} \overline{ \langle \,
Q^{\mu}_{N}(l)\,
  Q^{\mu+n}_{N}(l)\,\rangle} \,\,,\label{Cdef}
\end{equation}
in which $n=0,\dots, p-c-1$ and $C_{p-c}(l)=\Delta(l)$. The $p-c$
recursion relations for the $C_{n}$'s can be obtained in a
systematic way, as outlined in the appendix, leading to a variance
of the noise which depends on the spin-glass order parameter $q(l) =
\overline{\langle S(l) \rangle^2}$,
\begin{equation}
q(l)=\langle \int Dz\,\tanh^{2}\{\beta[\bxi.\bA\bm(l)+\Delta(l)
z]\}\rangle_{\bxi} \,\,\,. \label{qdef}
\end{equation}
In the case where $b=1$ (Hebbian noise), the variance follows the
simple form $\Delta^{2}(l+1)=\alpha + K^{2}(l)\Delta^{2}(l)$, where
$K(l)=\beta(1-q(l))$. In the case of the absence of stochastic noise
$(\alpha=0)$, the equations coincide with those for a recurrent
network with a parallel dynamics. The main difference between the
layered and the recurrent network is the absence of recurrent
connections in the former. The recurrent connections tend to amplify
the effects produced by the stochastic noise and the fact that the
equations are the same when $\alpha=0$ is simply due to the absence
of noise to amplify in the recurrent network in that case. In the
absence of stochastic noise the equations are also the same for the
recurrent network with either symmetric or asymmetric extreme
dilution \cite{AC01}.

Thus, the network dynamics is described by the recurrence
relations for the vector overlap (\ref{overlaprec}) and for all
the $C_{n}$'s that go into the variance of the noise. Although the
equations form an infinite set in the $p \rightarrow \infty$
limit, the number of significant $C_{n}$'s is finite making the
model solvable in practice. The transients and the dynamic
evolution of the network can be studied in full detail but here we
restrict ourselves to the stationary states.

\section{Results}

Consider first the ($T,\nu$) phase diagram of stationary states
for $\alpha=0$ and various numbers of condensed patterns $c$,
shown in Fig. 1(a). The Hopfield ansatz $m_{\mu}(1)=\delta_{\mu,
1}$, for $\mu=1,\dots, c$, is used as initial condition. The
states corresponding to fixed-point solutions and the phase
boundaries for $\nu>0.5$ are precisely the same as those found for
a recurrent network \cite{CT94}, since the equations for the order
parameters are exactly the same in the layered and in the
recurrent network when $\alpha=0$. The Hopfield-like states (H)
have one large condensed overlap component and the others are
either small or zero. The phase of symmetric-like states (S) has
equal or nearly equal overlap components, at high or low $T$,
respectively, ending at a paramagnetic phase (P), with $\bm= 0$
and $q= 0$. There is also a phase of non-trivially correlated
states (D) in which the correlation coefficients for the overlaps
with increasingly distant patterns from a stimulus become
gradually smaller, as will be seen below. The phase boundaries for
$\nu> 0.5$ are practically independent of $c$.
\begin{figure}[!h]
\begin{minipage}[c]{0.5\textwidth}
\includegraphics[width=7cm,height=6cm]{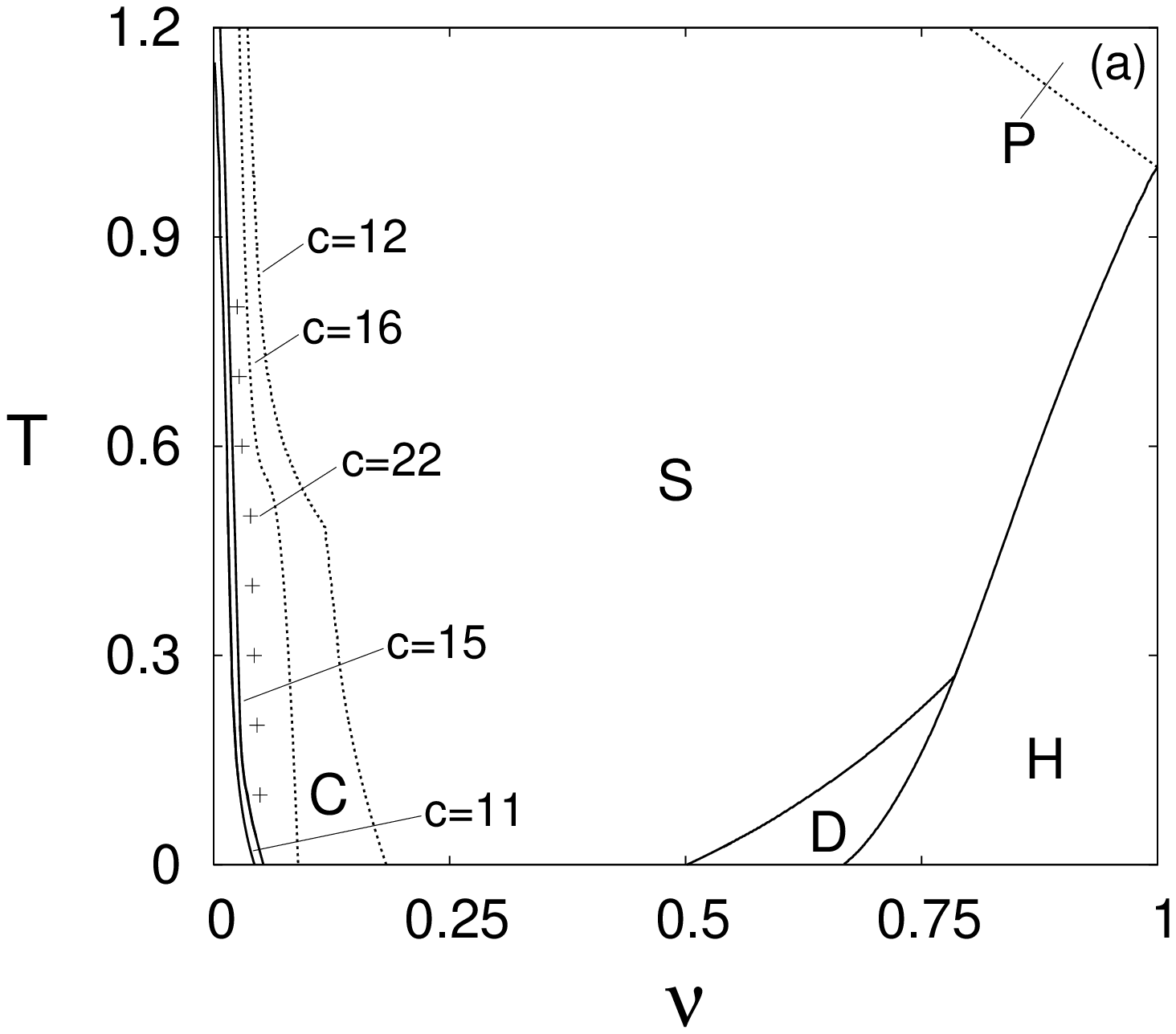}
\end{minipage}%
\vspace{1cm}
\begin{minipage}[c]{0.5\textwidth}
\includegraphics[width=7cm,height=6cm]{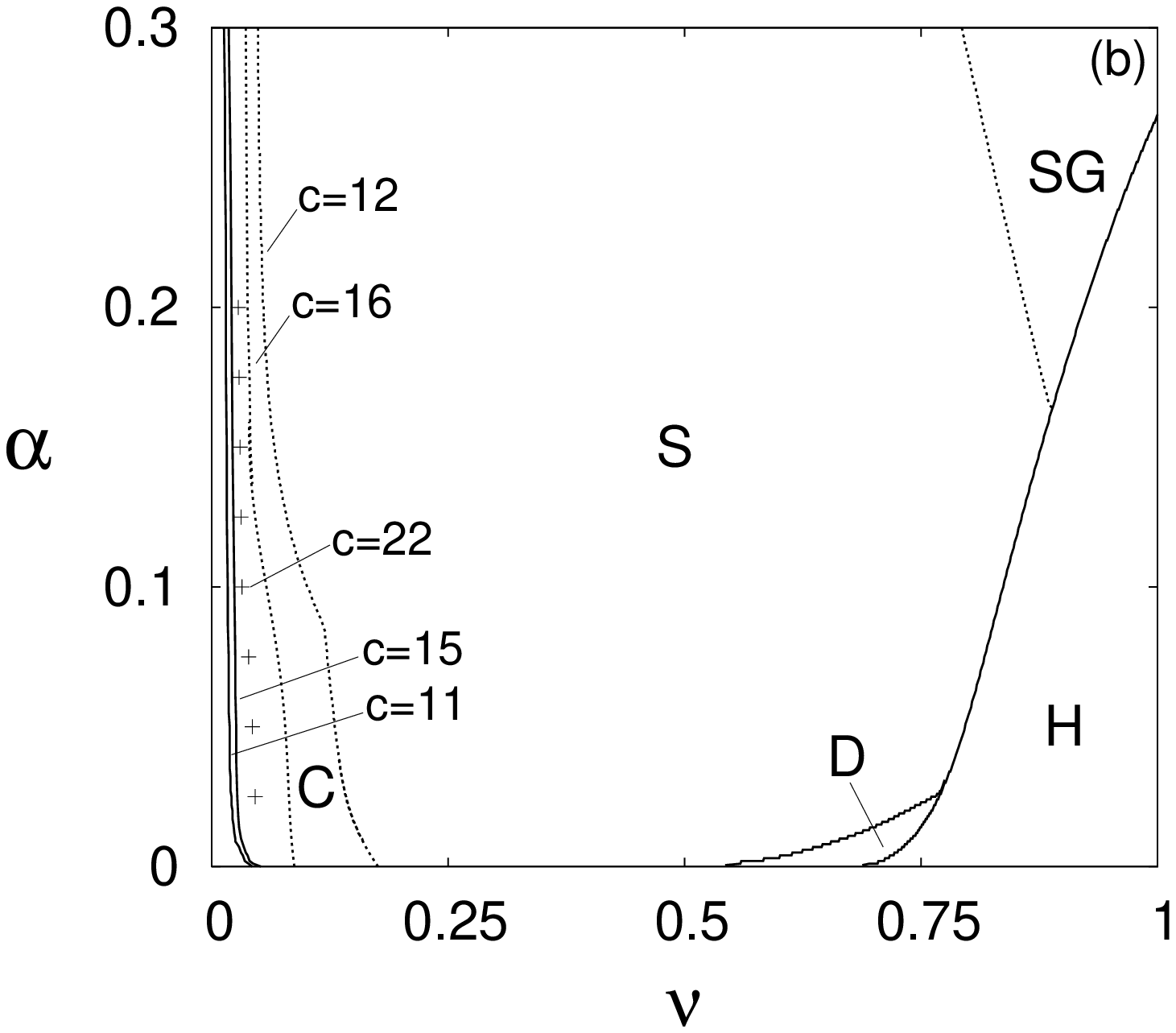}
\end{minipage}%
\caption{Phase diagrams (a) for $\alpha=0$ and (b) $T=0$ with a
purely Hebbian noise ($b=1$). The dotted and full lines indicate,
respectively, continuous and discontinuous transitions and the
phases are described in the text. Some points on the phase
boundary for $c=22$ are indicated with crosses.} \label{fig1}
\end{figure}

The novelty in the phase diagrams are stationary cyclic solutions
(C) of period two, for any $c$, in the region $0< \nu \lesssim 0.5$
in which $\bm(l+2)=\bm(l)$. These are the only stable states below
the phase boundaries for the presence of cyclic states. The main
difference with asymmetric sequence processing studied in an earlier
work \cite{MT05} is the presence in that case of cycles of period
$c$, as well as quasi-periodic states. Fig. 1 now shows that the
size of the cyclic phase decreases or increases with an increase of
$c$, if $c$ is even or odd, respectively, and there are no cyclic
solutions for $c<7$ in the latter case. These properties have been
checked by a linear stability analysis of the $S$ phase for
$\alpha=0$ in extension of earlier work \cite{CS92}. Clearly, the
periodic solutions are fairly robust to synaptic noise $T$. In
relation to a recent work \cite{MKO03}, we also studied our model
with a finite number of independent pairs of sequences for $\alpha =
0$ and low $\nu$ and found only cyclic solutions of period two, for
any number of stored sequence pairs and for any $c$ in each
sequence.

The effects of stochastic noise due to a macroscopic number of
patterns $p=\alpha N$ are shown in Fig. 1(b) for $T=0$, $b=1$
(Hebbian noise) and various $c$. For the non-condensed overlaps we
choose the initial $C_{n}^2(1)=\alpha$, for all $n$, and for the
condensed overlaps we take again a Hopfield initial condition. As
usual in the layered network, there is now a spin-glass phase with
$q\neq 0$ (labeled SG). For $\nu=1$ we recover the critical storage
ratio $\alpha_{c} \simeq 0.269$ for the Hebbian layered network
model \cite{DKM89}. Also here we find the same kind of stationary
states as in Fig. 1(a) for $\nu > 0.5$ and stationary cyclic states
of period two in the region of low $\nu$, with the absence of the
latter for $c<7$ in the case of odd $c$. Again, the boundaries
between phases of fixed-point states are fairly independent of $c$,
while the cyclic phase boundaries have a similar dependence on $c$
as in the $(T,\nu)$ phase diagram. We also found that the cyclic
phase boundaries almost do not vary beyond $c=11$ and $c=22$, for
odd and even $c$, respectively. They are very close and they should
be independent of $c$ in the large $c$ limit.
\begin{figure}[!h]
\center
\includegraphics[width=7cm,height=6cm]{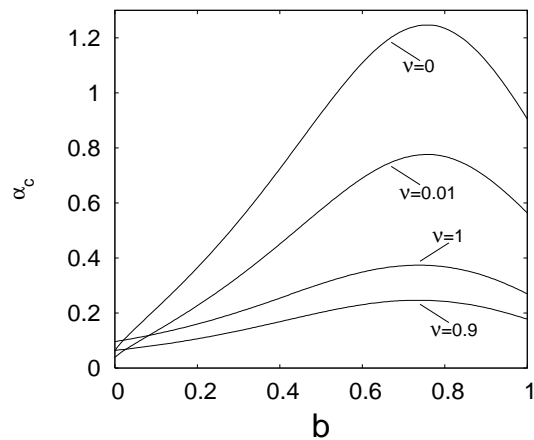}
\caption{Critical storage ratio as a function of $b$ for $T=0$ and
$c=13$ in the Hopfield-like phase ($\nu=1$ and $\nu = 0.9$) and in
the cyclic phase ($\nu=0$ and $\nu = 0.01$).}
\end{figure}

Although Fig. 1 gives a fair account of the phase diagram for
Hebbian noise ($b=1$), one may ask which is the effect of other
noise parameters ($b\neq 1$) and in Fig. 2 we show the critical
values $\alpha_c$ for the existence of two typical states in each of
the phases $H$ and $C$ as functions of $b$, for $c=13$ and $T=0$. In
the case of fixed-point states, that is within the phase $H$ (and
also for the $D$ phase, not shown in the figure), $\alpha_c$
increases with increasing $\nu$ for a given $b$ whereas in the case
of cyclic states $\alpha_c$ decreases, as one would expect. Similar
results are obtained for $c=12$. There is a maximum $\alpha_c$ for
an optimal $b\simeq 0.748$ and the kind of noise becomes more
relevant for a dominant sequential part (small $\nu$) in the signal
of the local field.

We consider next the solutions for the stationary overlaps that
describe the long-time behavior of the network, for a weak Hebbian
term. The overlaps for states in the phase diagrams bifurcate from a
fixed-point behavior in the $S$ phase to stable stationary limit
cycles on the continuous (discontinuous) transition from the S to
the C phase, for even (odd) $c$, respectively. In order to describe
one of our main results, that is the nature of the cyclic behavior
we illustrate this for the overlaps in Fig. 3 for $c=13$ when
$\alpha=0$ for a typical low synaptic noise of $T=0.3$. The
bifurcation diagram contains the stationary values of the first
seven overlaps as functions of $\nu$, with $m_{\mu}(1) =
\delta_{\mu,1}$ as an initial condition. Each overlap assumes a
larger and a smaller value in the cyclic phase marked with the same
symbols (for clarity, only the larger one is labeled), with a
decreasing oscillation amplitude as we move away from the stimulated
pattern. For quite higher noise levels, say $T=1.25$ and appropriate
values of $\nu$ (see below), all the pairs of overlap components
keep oscillating between the same upper and lower values. Thus, for
such high levels of $T$, the cyclic phase is already a phase of a
pair of symmetric states, one with all equal larger condensed
overlaps and the other with all equal smaller overlaps. In addition,
the overlap components have the symmetry $m_{\mu + n}(t) = m_{\mu -
n}(t)$, where $\mu$ is the stimulated pattern and
$n=1,\dots,(c-1)/2$.
\begin{figure}[!h]
\center
\includegraphics[width=7cm,height=6cm]{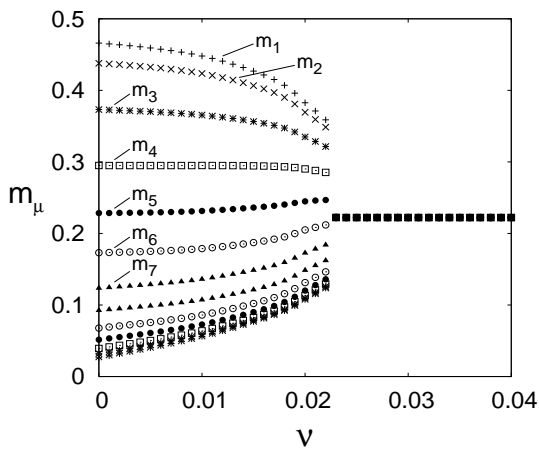}
\caption{Overlaps for the stationary cycles of period two, discussed
in the text, that bifurcate from the symmetric phase for $c=13$,
$\alpha=0$ and $T=0.3$.}
\end{figure}

In the case of even $c$, there is first a continuous transition to
a pair of overlaps with decreasing $\nu$ such that all solutions
keep oscillating between an upper and a lower value. There is a
further, discontinuous transition, for lower values of $\nu$, to
distinct pairs of overlaps that is similar to the discontinuous
transition for the case of odd $c$, with the same symmetry between
overlaps now for $n=1,\dots,(c-2)/2$. Thus, the behavior of the
overlap components has a rich structure which depends on whether
$c$ is even or odd.

In order to demonstrate that the cyclic states may be non-trivial
correlated attractors for low $T$ and $\nu$, we consider next the
correlation coefficients between the attractors corresponding to
any two condensed patterns, $\xi^{\mu}$ and $\xi^{\nu}$ a distance
$d=|\mu - \nu|$ away, defined here as
\begin{eqnarray}
C_d &=& \frac{1}{C} \sum_{i} (\langle\sigma_{i}^{\mu} \rangle -
\overline{\langle\sigma_{i}^{\mu}\rangle}\,) (\langle
\sigma_{i}^{\nu}\rangle-
\overline{\langle\sigma_{i}^{\nu}\rangle}\,) \nonumber \\
&=& \frac{\langle \tanh(\beta h^{\mu}) \tanh(\beta h^{\nu})
\rangle_{\bxi}}{\langle \tanh^{2}(\beta h^{\mu})
\rangle_{\bxi}}\,\,\,, \label{11}
\end{eqnarray}
where $\langle\sigma_{i}^{\rho}\rangle$ is the attractor
corresponding to the initially stimulated pattern $\rho$ and
$\overline{\langle\sigma_{i}^{\rho}\rangle}$ is its mean value over
the network which is zero for the unbiased patterns we are using
here.

We show in Fig. 4 the dependence of the correlation coefficients
for the states corresponding to the larger overlap component of
each pair of states with the distance $d$ to increasingly distant
condensed patterns in the sequence from a given stimulated one,
for $c=13$ and $\alpha=0$, that is, in the absence of stochastic
noise where the results are the same for the layered and the
recurrent network. We do this, as indicated, for
$(T,\nu)=(0.3,0.01)$ and also for $(1.25,0.001)$ both within the
phase of cyclic states.
\begin{figure}[!h]
\center
\includegraphics[width=7cm,height=6cm]{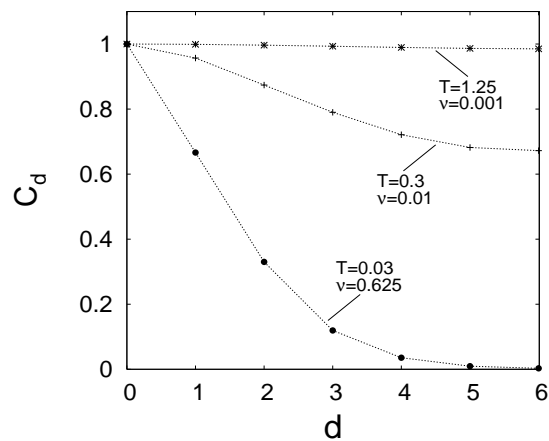}
\caption{Correlation coefficients between attractors as a function
of the distance $d$ from a reference pattern, defined in the text,
for $c=13$ and $\alpha=0$ in the phases $C$ (two upper curves) and
$D$ (lower curve). The lines are a guide to the eye.}
\end{figure}
In the first case, where each overlap component assumes distinct
larger and smaller values, there are also correlation coefficients
for the smaller values, not shown in the figure, which turn out to
decrease less rapidly than the coefficients for the larger values.
The reason for this is that there is a weaker distinction between
the smaller components than among the larger ones. For the higher
$T=1.25$ the correlation coefficients are already independent of $d$
due to the fact that all the pairs of overlap components are
oscillating between the same upper and lower values.

In distinction to the fixed-point correlated states in phase D,
shown for our model by the lower set of results in the figure for
$T=0.03$ and $\nu=0.625$, the correlation coefficients for the
cyclic states do not decrease to zero and, instead, exhibit a
behavior typical of quasi-symmetric states for low $T$.

\section{Discussion}

To summarize our results, we obtained a closed-form attractor
dynamics for a feed-forward layered network model of binary units
and patterns in terms of a finite number of macroscopic order
parameters for the competition between pattern reconstruction and
symmetric sequence processing. The dynamics is a parallel one, in
which all units in each layer are updated simultaneously, and the
work presented here is restricted to the stationary states of the
dynamics which exhibits either fixed-point or cyclic behavior of
period two, depending on the relative strength of the
interactions. Either kind of behavior is an emergent property of
the network which is an outcome of the dynamics.

Full phase diagrams of stationary network behavior were obtained,
either for a finite loading of patterns or in the saturation limit
for the storage of a macroscopic number of patterns. In the case of
a balanced or dominant Hebbian term, that is for $\nu \geq 0.5$, we
obtained qualitatively the same phase diagrams as those found in
work by previous authors for a recurrent attractor network
exhibiting Hopfield-like states, correlated states and symmetric
ordered states \cite{CT94}. This suggests that the layered-network
dynamics discussed here may also serve to make qualitative
predictions about a recurrent network in the case of a weak Hebbian
term. Of course, to get the right quantitative behavior expected for
a recurrent network in the storage saturation limit, and for
detailed comparison with experiments, one has to start by including
lateral connections between units in a layer but this involves a
more complicated dynamics which is beyond the scope of this paper.
For $\nu$ below $0.5$ we have first a regime of symmetric
fixed-point behavior and for smaller values of $\nu$ we find the
cyclic attractors discussed in this work. These are states that have
decreasing correlation coefficients with increasingly distant
attractors from an initially stimulated pattern for a sizeable range
of synaptic noise.

The stationary overlaps between the states of the network and the
condensed patterns describe the long-time behavior of the system. As
we saw, the overlaps for states within the cyclic phase are always
periodic with period two, oscillating either between a distinct pair
of upper and lower values for each overlap component or between the
same pair of values for all components, depending on the parity of
$c$ and on the state in the $(\alpha,T,\nu)$ phase diagram. We argue
that the cyclic behavior of period two is a property that follows
essentially from the nature of the interactions, that is, from the
strong symmetric sequential term rather than being an artifact of
the model due to the lack of lateral connections between the units
or having binary units. First, in support of our claim, work in
progress on the parallel dynamics of a fully connected recurrent
network of binary units and patterns indicates, indeed, that there
are exclusively cyclic states of period two in a finite region of
the phase diagram \cite{MT06}. Second, we checked explicitly that
numerical simulations with threshold-linear units \cite{Br94} in our
feed-forward layered network yield only cyclic states of period two
for typical low values of $\alpha$ and $\nu$, at $T=0$. The specific
regions of the phase diagram where distinct pairs of upper and lower
values or the same pair of values for the overlaps appear depends,
of course, on the details of the model.

The phase diagrams found in this work, are quite different from
those for {\it asymmetric} sequence processing \cite{CS92,MT05}.
Indeed, the latter exhibit a $\nu \Leftrightarrow (1-\nu)$ duality
that appears in the form of symmetric phase diagrams with a
correspondence between fixed-point solutions for large $\nu$ and
stationary cycles of period $c$ for small $\nu$. There is,
apparently, no such duality in the case of symmetric sequence
processing and whenever stable cycles appear they are of period two,
independently of the number of condensed patterns $c$. Also, the
strong $c$ dependence of the stability of the cyclic phase is in
contrast with the results for asymmetric sequence processing, in
which the boundary of the cyclic phase practically does not depend
on the number of condensed patterns \cite{CS92,MT05}.

Although the work presented here is restricted to a layered
feed-forward network with no lateral interactions, it is expected
to exhibit further features of a recurrent network beyond those
pointed out above, in particular, the robustness to both synaptic
($T$) and to stochastic noise ($\alpha$) due to the non-condensed
patterns over a sizeable ratio $\nu/(1-\nu)$ of the relative
strength of the Hebbian to sequential interaction (Figs. 1(a) and
1(b)), respectively. The robustness becomes stronger in the case
of decreasing even values of the number of condensed patterns $c$
and weaker for odd values of decreasing $c$.

The model used here has several limitations with respect to a closer
to biological network which do not allow to make the proper
quantitative predictions to compare with experiments, mainly the
binary full activity units, in place of continuous or integrate and
fire neurons, and unbiased (high activity) patterns, without lateral
interactions between units in the same layer. Despite those
limitations, there is the possibility of making extended qualitative
predictions for the kind of visual-task experiments in the IT
primate cortex and their interpretation in terms of correlated
states in simple models for a recurrent network [8-12]. Those works
provided a connection between a fixed-point attractor dynamics in a
recurrent network and persistent activity in a biological system.
Our work suggests a connection between a periodic attractor dynamics
in a recurrent network trained with patterns in a random order and
with patterns in a sequence and a kind of oscillating persistent
activity in the IT cortex with the specific cyclic behavior
discussed in this work. The original experiments were based on
intensive training with patterns in a random order and with patterns
in a sequence. One may argue that {\it if} the training of a primate
with visual patterns in random order, which is supposed to be a
realization of a Hebbian rule, is not sufficiently strong, one may
have a situation as that described here for small $\nu$ with the
presence of correlated cyclic states of period two, in which the
correlation coefficients decay with increasing distance from the
stimulus, up to a finite value.

The results presented here should stimulate further theoretical and
experimental work for the case of weak Hebbian reenforcement of
patterns. It may also lead to interesting
applications in information processing in networks \cite{MA95}.\\

{\bf Acknowledgements}\\

The work of one of the authors (WKT) was financially supported, in
part, by CNPq (Conselho Nacional de Desenvolvimento Cient\'{\i}fico
e Tecnol\'ogico), Brazil. Grants from CNPq and FAPERGS
(Funda\c{c}\~ao de Amparo \`a Pesquisa do Estado de Rio Grande do
Sul), Brazil, to the same author are gratefully
acknowledged. F. L. Metz acknowledges a fellowship from CNPq.\\

\appendix*
\section{Recursion relations}

We present next an outline of the derivation of the recursion
relations for all $C_{n}$'s. It is based on an extension of the
usual procedure \cite{MT05,DKM89} adapted to our specific synaptic
matrix. The main step is the recursion relation for the
correlation between any two microscopic overlaps with the
non-condensed patterns which enter in Eq.(9).

We start with the definition of the average
\begin{equation}
\overline{\langle\,M^{\prime\mu}_{N}\,M^{\prime\nu}_{N}\,\rangle}
=\frac{1}{N^2}\sum_{ij}\,\overline{\xi_{i}^{\prime\mu}\,
\xi_{j}^{\prime\nu}\langle S^{\prime}_{i}
S^{\prime}_{j}\rangle}\,\,, \label{A1}
\end{equation}
for $\mu, \nu = c+1,\dots,p$, in which the primed (unprimed)
variables refer to layer $l+1$ ($l$), respectively. The averages are
explicitly calculated only with respect to primed variables since
the averages over the underlying unprimed variables in the lower
layer are taken care by means of the law of large numbers. Writing
Eq.(A1) in the form
\begin{equation}
\overline{\langle\,M^{\prime\mu}_{N}\,M^{\prime\nu}_{N}\,\rangle}
=\frac{1}{N^{2}}\sum_{i}\overline{\xi_{i}^{\prime\mu}\,\xi_{i}^{\prime\nu}}
+ \frac{1}{N^{2}}\sum_{i \neq
j}\overline{\xi_{i}^{\prime\mu}\,\xi_{j}^{\prime\nu}\langle
  S^{\prime}_{i}\rangle \langle S^{\prime}_{j}\rangle}\,\,,
\label{A2}
\end{equation}
allows us to take the thermal averages leaving, in our case of
binary patterns,
\begin{equation}
\overline{ \langle\,M^{\prime\mu}_{N}\,M^{\prime\nu}_{N}\,\rangle}
= \frac{\delta_{\mu\nu}}{N} + \frac{1}{N^{2}} \sum_{i \neq j}
\overline{\tanh(\beta
\xi_{i}^{\prime\mu}h_{i}^{\prime})\,\tanh(\beta
\xi_{j}^{\prime\nu}h_{j}^{\prime})}\,\,, \label{A3}
\end{equation}
using the fact that the patterns are uncorrelated variables with
$\overline{\xi_{i}^{\prime\mu}\,\xi_{i}^{\prime\nu}} =
\delta_{\mu\nu}$. The embedding field
$\xi_{i}^{\prime\mu}h_{i}^{\prime}$ is given by
\begin{equation}
\xi_{i}^{\prime\mu}h_{i}^{\prime} =
\xi_{i}^{\prime\mu}(\bxi^{\prime}.\bA\bm) + Q^{\mu}_{N} +
\xi_{i}^{\prime\mu}\omega^{\prime}_{i}\,\,, \label{A4}
\end{equation}
where $\{\omega^{\prime}_{i}\}$ is a set of independent Gaussian
random variables $\omega^{\prime}_{i} = \sum_{\rho \neq \mu}^{p}
\xi_{i}^{\prime\rho} Q^{\rho}$ with mean
$\overline{\omega^{\prime}_{i}}=0$ and width $\Delta$ defined by Eq.
(6). Incidentally, the same expression for the embedding field
without the need of separating a single term on the right may be
used to obtain the recursion relation for the overlap components
that yield Eq. (8).

Turning now to the configurational average in Eq. (A3) over primed
variables, it decouples into a product of averages. Making an
expansion to leading order in $Q^{\mu}_{N} = O(1/\sqrt{N})$ we
obtain
\begin{eqnarray}
 &\overline{\tanh(\beta \xi_{i}^{\prime\mu}h_{i}^{\prime})} =  \overline{\tanh
 \beta( \xi_{i}^{\prime\mu}(\bxi^{\prime}.\bA\bm) +
 \xi_{i}^{\prime\mu}z^{\prime}_{i})}& \nonumber\\
 &+ \beta  Q^{\mu}_{N} \overline{[1 - \tanh^{2}
 \beta( \xi_{i}^{\prime\mu}(\bxi^{\prime}.\bA\bm) +
 \xi_{i}^{\prime\mu}z^{\prime}_{i})]\,\,.}&
\label{A5}
\end{eqnarray}
Averaging first with respect to the variable $\xi_{i}^{\prime\mu}$
and then taking the configurational average over the Gaussian
variable we obtain
\begin{equation}
\overline{\tanh(\beta \xi_{i}^{\prime\mu}h_{i}^{\prime})} =
K\,Q^{\mu}_{N}\,\,, \label{A6}
\end{equation}
in which $K=\beta\,(1-q)$, with $q$ defined in Eq. (10).
Substituting Eq. (A6) in (A3) we obtain, to $O(1/N)$,
\begin{equation}
  \overline{\langle\,M^{\prime\mu}_{N}\,M^{\prime\nu}_{N}\,\rangle}
= \frac{\delta_{\mu\nu}}{N} + K^{2}\,Q^{\mu}_{N}\,Q^{\nu}_{N}\,\,.
\label{A7}
\end{equation}

We can now apply Eq. (9) to layer $l+1$ and write it in terms of the
correlations
$\overline{\langle\,M^{\prime\,\mu}_{N}\,M^{\prime\,\nu}_{N}\,\rangle}$
with the aid of Eq. (7). Then, Eq. (A7) yields the $p-c$ recursion
relations\\
\begin{eqnarray}
 \Delta^{\prime\,2}&=& b^2[\,\alpha + K^2 \Delta^2\,] + 4b(1-b) K^2 C_{1}^2 \nonumber \\
&+&  2(1-b)^2  K^2[\,\alpha/K^2 + \Delta^2 + C_{2}^2\,]\,,  \label{A8}   \\
 C_{1}^{\,\prime\,2}&=& b^2 K^2 C_{1}^{2} + 2b(1-b) K^2 [\alpha/K^2 +
 \Delta^2 \nonumber \\
 &+& C_{2}^2]+ (1-b)^{2}\,K^2[\,3\,C_{1}^2+C_{3}^2]\,,   \label{A9} \\
 C_{2}^{\,\prime\,2}&=& b^2\,K^2 C_{2}^{2} + 2b(1-b) K^2[\,C_{1}^{2}+C_{3}^2] \nonumber \\
&+& (1-b)^{2} K^2 [\,\alpha/K^2 + \Delta^2 \nonumber \\
&+& 2 C_{2}^2 + C_{4}^2\,]\,, \label{A10}
\end{eqnarray}
up to
\begin{eqnarray}
C_{n}^{\,\prime\,2}&=& b^2\,K^2 C_{n}^{2} + 2b(1-b) K^2[\,C_{n-1}^2+C_{n+1}^2]  \nonumber\\
&+&  (1-b)^{2} K^2 [\,C_{n-2}^{2} + 2 C_{n}^2 + C_{n+2}^2\,]\,,
\label{A11}
\end{eqnarray}
in which $n=3,4,\dots,p-c-1$, and these equations have to be solved
numerically.

The extension to $Q$-state neurons and patterns, for $Q\geq 3$ is
straightforward, as well as for continuous neurons.

\end{document}